\documentstyle[10pt,epsf]{iopart}                % use this for journal style
\newcommand{\insertplot}[1]{
\begin{center}\leavevmode\epsfysize=5.1cm \epsfbox{#1}\end{center}}

\newcommand{\insertplotg}[1]{
\begin{center}\leavevmode\epsfysize=7.7cm \epsfbox{#1}\end{center}}

\def\Journal#1#2#3#4#5{#1 {\it #2} #3 {\bf #4} #5}
\def\Preprint#1#2{#1 {\it Preprint} #2}

\def\PL{Phys. Lett.}
\def\PR{Phys. Rev.}
\def\PRL{Phys. Rev. Lett.}

\begin{document}

%%%%%%%%%%%%%%%%%%%%%%% TITLE PAGE %%%%%%%%%%%%%%%%%%%%%%%%%%%%%%

\title{The production of charm mesons from quark matter
at CERN SPS and RHIC }

\author{P. L\'evai, T.S. Bir\'o, P. Csizmadia, T. Cs\"org\H o and J. Zim\'anyi}

\address{
KFKI Research Institute for Particle and Nuclear Physics, \\ 
	 P. O. Box 49, Budapest, 1525, Hungary }

\begin{abstract}
We study the production of charm mesons and other charm baryons
from quark matter at 
CERN SPS and RHIC energies. Using quark coalescence models 
as hadronization mechanism, we predict
particle ratios, absolute yields and transverse momentum spectra.
\end{abstract}

\vspace*{-0.5 cm}


%%%%%%%%%%%%%%%%%%%% INTRODUCTION %%%%%%%%%%%%%%%%%%%%%%%%%%%%%%%%%

\section{Introduction}
 
During the last decade $J/\psi$ production was investigated with 
great interest at the CERN Super Proton Synchrotron (SPS), 
and an anomalous suppression has been detected in Pb+Pb collisions
at 158 GeV/nucleon bombarding energy \cite{NA50supr}. This
result is is assumed to be the strongest evidence of QGP formation
(for an overview see Ref.~\cite{Satz00}).
However, the unambiguous confirmation of the QGP formation requires more
charm related data, e.g. absolute yields, transverse momentum spectra and 
the characteristics of charm hadron production from deconfined matter.

In this paper we summarize our results obtained by 
assuming quark coalescence as microscopical hadronization mechanism
of the possibly produced deconfined phase at CERN SPS energy.
The ALCOR \cite{ALCOR,Transchem,ALCORc} and MICOR \cite{MICOR,MICORc} 
hadronization models were extended by including charm quark and
the consecutive charm hadron formation.

\section{The number of charm quarks in heavy ion collision at SPS energy}

Perturbative QCD calculations predict  small number of newly 
produced charm quark-antiquark pairs in the Pb+Pb collision at CERN SPS energy,
$N_{c{\overline c}} \approx 0.1$. On the other hand statistical models
obtain a much larger number, $N_{c{\overline c}} \approx 8.5$ 
\cite{Gazd99}. If we assume that deconfinement appears after
the formation of a hot dense string matter where strings become ropes
in the presence of large density 
(characterized by an increased string constant, $\kappa$),
then quark-antiquark pair creation will
be dominated by the breaking of these in-medium strings/ropes.
The enhancement of the heavy quark-antiquark production,
$\gamma_X$, can be calculated
by means of  the Schwinger mechanism \cite{Blei00}:
\begin{equation}
\gamma_X = \frac{P(X{\overline X})}{P(q{\overline q})} =
\exp \left( - \frac{\pi (m_X^2 - m_q^2)}{\kappa} \right)
\end{equation}
If we increase the usual value of the string constant
$\kappa=1.0$ GeV/fm with 10 \%, then an enhancement factor 24 appears for
charm quarks with mass $m_c=1.5$ GeV ($m_q=223$ MeV).
20 \% increase in $\kappa$ would generate an enhancement factor 350 !!! \\
Thus relatively small change in the string breaking could modify 
dramatically the total charm production in heavy ion collisions.
If relatively large number of charm quarks can be produced in the initial
phase of the heavy ion collision via in-medium breaking of the primary 
strings, then after deconfinement the quark coalescence can become 
a reliable description of charm hadron formation from the plasma state.

\section{Quark coalescence}

In our description quark coalescence is the basic hadronization mechanism.
It is assumed that the number of directly produced hadrons is given by 
the product of the 
number of quarks (or anti-quarks) from which those hadrons are
produced, multiplied by coalescence coefficients $C_h$ and
by non-linear normalization coefficients $b_q$, that take into account
conservation of quark numbers during quark coalescence 
\cite{ALCOR,Transchem}.
The number of various quarks is denoted by the
symbol usual for that type of particles, e.q. $u$, $d$, $s$ and $c$
denote the number of light, strange and charmed quarks, respectively.
The following examples display the appropriate coalescence relations
(see Ref.~\cite{ALCORc}):
\begin{eqnarray}
p^+\,(uud)&=
C_p \cdot (b_u\, u)^2 \cdot (b_d\, d) \hskip 1.10 truecm
K^+ \, (u \overline s) \ &=
C_K \cdot (b_u \, u) \cdot (b_{\overline s} \, {\overline s}) \nonumber \\
& \ ...   \hskip 1.00 truecm  & \ ...  \nonumber \\
\Omega\,(sss) &= 
C_{\Omega} \cdot (b_s\, s)^3 \hskip 2.50 truecm
\phi \, (s \overline s) \  &=
C_{\phi}
\cdot (b_s \, s) \cdot (b_{\overline s} \, {\overline s}) \nonumber \\
\Omega_c\,(ssc) &= 
C_{\Omega}^c \cdot (b_s\, s)^2 \cdot (b_c\, c) \hskip 1.10 truecm
D^+ \, (c {\overline d} ) \  &=
C_D \cdot  (b_c \, c) \cdot (b_{\overline d} \, {\overline d}) \nonumber \\
\Omega_{cc}\,(scc) &= 
C_{\Omega}^{cc} \cdot (b_s\, s) \cdot (b_c\, c)^2 \hskip 1.10 truecm
D_s^+ \, ( c \overline s )   &=
C_D^s \cdot (b_c \, c) \cdot (b_{\overline s} \, {\overline s}) \nonumber \\
\Omega_{ccc} \,(ccc) &= 
C_{\Omega}^{ccc} \cdot (b_c\, c)^3 \hskip 1.95 truecm
J/\psi \, (c\overline c)   &=
C_{J/\psi} \cdot (b_c \, c) \cdot (b_{\overline c} \, {\overline c})
\nonumber 
\end{eqnarray}
The non-linear coalescence factors
$b_u$, $b_d$, $b_s$, $b_c$ and the
$b_{\overline u}$, $b_{\overline d}$, $b_{\overline s}$, $b_{\overline c}$
are determined unambiguously from the requirement
that the number of constituent quarks and anti-quarks do not  change
during the hadronization, and that all initially available
quarks and anti-quarks
have to end up in the directly produced hadrons.
This constraint is a basic assumption in all models of quark coalescence.
The correct quark counting yields to 8 coupled non-linear
equations, expressing the conservation of the number of quarks
and antiquarks.
Without solving these non-linear equations, one can recognize 
interesting relations among the different hadron ratios.
For example certain meson ratios play the role
of step functions among different antibaryon to baryon ratios. This is
illustrated in Figure 1 from Ref.~\cite{ALCORc},
where the $u$ and $d$ quarks are considered
together as $q$ light quark.

\insertplot{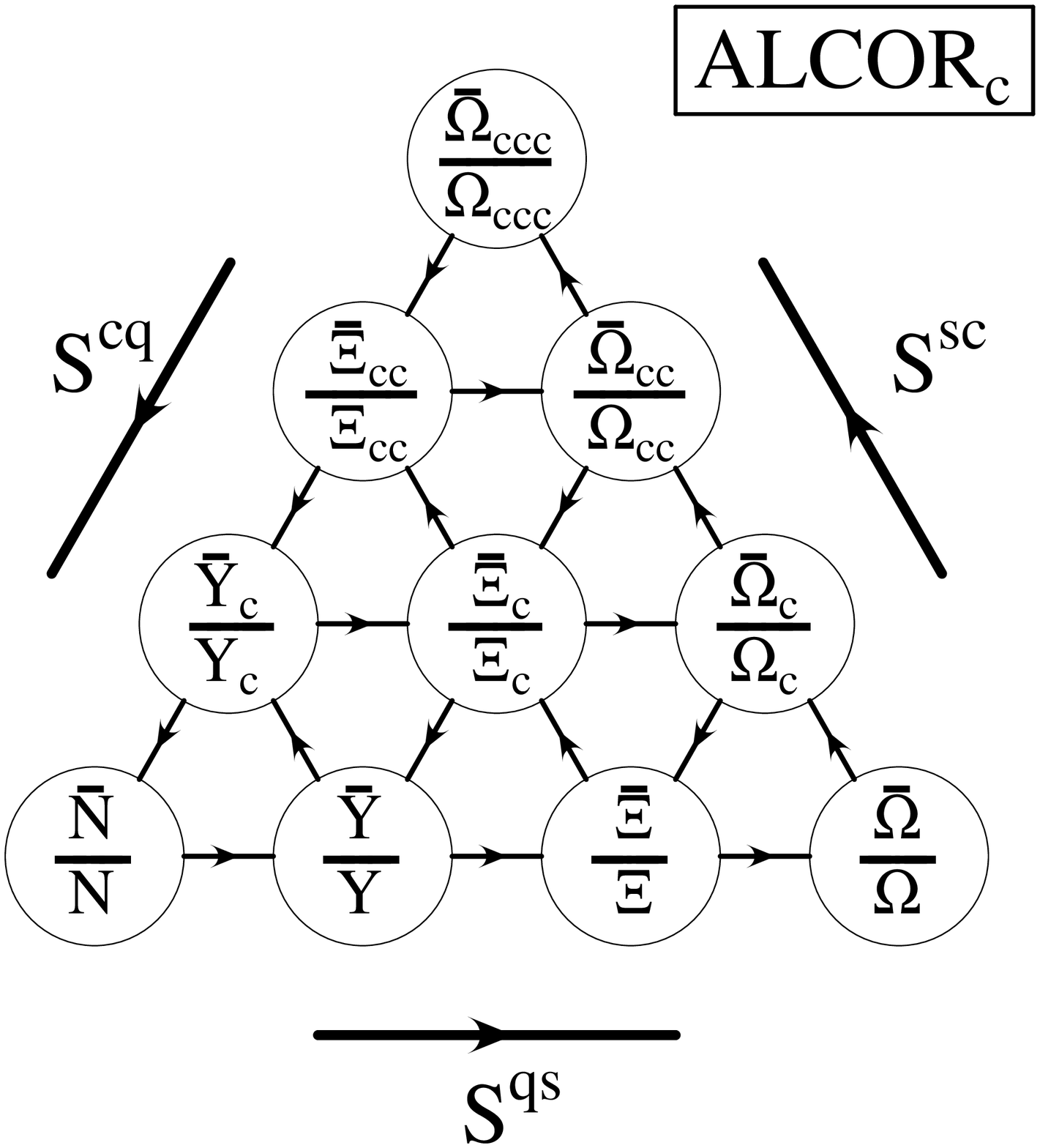}
 \begin{center}
  \begin{minipage}[t]{13.054cm}
  { {\bf Figure 1.}
Mesonic step factors ${\cal S}^{qs}$, ${\cal S}^{sc}$,
${\cal S}^{cq}$ and  antibaryon to baryon ratios. }
  \end{minipage}
 \end{center}

\newpage

Here the step factors are defined as:
\begin{eqnarray}
{\cal S}^{qs} \equiv \ 
\frac{K\,(q\overline s)}{{\overline K}\, ({\overline q} s )} &=&
\left[ \frac{b_q\, q}{b_{\overline q} \, {\overline q}} \right] \cdot
\left[ \frac{b_{\overline s} \, {\overline s}}{b_s\, s} \right] \ , \\
{\cal S}^{sc}  \equiv \
\frac{{\overline D}_s\,(s\overline c)}{ D_s\, ({\overline s} c )} &=&
\left[ \frac{b_s\, s}{b_{\overline s} \, {\overline s}} \right] \cdot
\left[ \frac{b_{\overline c} \, {\overline c}}{b_c\, c} \right] \ ,  \\
{\cal S}^{cq}  \equiv \
\frac{ D\, ({\overline q} c )}{{\overline D}\,(q\overline c)} &=&
\left[ \frac{b_c\, c}{b_{\overline c} \, {\overline c}} \right] \cdot
\left[ \frac{b_{\overline q} \, {\overline q}}{b_q\, q} \right] \ .
\end{eqnarray}
These properties lead to the identity
${\cal S}^{qs} \cdot {\cal S}^{sc} \cdot {\cal S}^{cq} \equiv  1 $,
which can be rewritten as a measurable relation between these mesonic ratios:
$({{\overline D_s} / D_s})/({{\overline D} / D}) =
{\overline K}/{K} $.
These relations are valid at both CERN SPS and RHIC energies, if quark
coalescence describes the hadronization of the produced deconfined phase.

\section{Predictions for charm hadron yields at SPS and RHIC }

Total hadron yields can be determined through the numerical
solution of the appropriate 8 non-linear equations of quark 
number conservation. The input values are the newly produced
quark-antiquark numbers, $N_{u{\overline u}}$, $N_{d{\overline d}}$,
$N_{s{\overline s}}$ and $N_{c{\overline c}}$.
At CERN SPS energy the earlier ALCOR  calculations 
could successfully reproduce most of the
experimental data within errorbar
(even the ${\overline \Omega}/\Omega$ ratio) \cite{Transchem}. 
%The input values for
%the newly produced quark-antiquark pairs were the followings:
%$N_{u{\overline u}} = N_{d{\overline d}}=383$, $N_{s{\overline s}} = 215$.
%These numbers were fixed from the values $h^- = 680$ and $K^+=96$.

Including the charm quark into this model, one experimental data
about charm hadron production is needed to determine the full
charm production yield. We choose the known ratio $J/\psi / h^-  =
1.0 \cdot 10^{-6}$ \cite{Gazd99}. The ALCOR model can reproduce this
data with  $N_{c{\overline c}}=3.4$ \cite{ALCOR4}. 
The obtained charm hadron yields (see Table 1) 
display a factor of 12 enhancement relative to the
perturbative QCD predictions, but they reach only 40 \%
of the predictions from the statistical model \cite{Gazd99}.

We repeat our calculation at RHIC energies (see Table 1), assuming that the 
relative strange and charm quark-antiquark pair production remains 
the same, only the absolute yields are scaled with increasing energy,
following the increase in the number of $h^-$ ($h^-$ was obtained
from the HIJING model). Table 1 shows that detectable numbers of
charm hadrons ($D$, $\Lambda_c$, ${\overline \Lambda}_c$) 
ought to be  produced at RHIC.

\section{Transverse momentum spectra of the D and $J/\psi$ mesons}

The transverse momentum spectra of the produced hadrons can be calculated
by the MICOR model \cite{MICOR,MICORc}. Applying this model 
one can determine the hadronization temperature $T_{HAD}$ and the transverse
flow of the hadronizing quarks. At CERN SPS energy 
$T_{HAD}=175$ MeV and $v_T=0.46 \pm 0.05$ 
fit the data in the Pb+Pb collision \cite{MICOR}. 
Including charm quarks into this model
we can determine the transverse momentum spectra of the 
$D$ and the $J/\psi$ mesons. Figure 2 displays the obtained values
$T_{eff}^D=259 \pm 19$ MeV and $T_{eff}^{J/\psi} = 315 \pm 35$ MeV
for the slope parameter of the transverse momentum spectra \cite{MICORc}.
We urge future experiments
to measure these quantities at CERN SPS.

%%%%%%%%%%%%%%%%%%% ACKNOWLEDGMENT %%%%%%%%%%%%%%%%%%

\ack
%\section*{Acknowledgments}
%\acknowledgments

This work was supported by the Hungarian OTKA Grant T024094, 
T025579, and partly by
the US-Hungarian Joint Fund No. 652.
\newpage

\begin{table}
\caption{
Charm hadron production at CERN SPS and RHIC energies
from the ALCOR model}
\begin{indented}
\item[]\begin{tabular}{@{}lllll}
\br
            & $\sqrt{s}=17.4$ A GeV
& $\sqrt{s}=56$ A GeV  & $\sqrt{s}=130$ A GeV  \\
\br
 $h^-$ 
& 680     &1131     &1828   \\
\br
 $N_{c{\overline c}} $  
& 3.4     & 6.7     & 12    \\
 $\langle D^0 + {\overline D}^0 \rangle $  
&2.4      &4.2      &6.4    \\
 $\langle J/\psi \rangle / h^- $
&$1.0 \cdot 10^{-6}$ & $1.5 \cdot 10^{-6}$ & $1.6 \cdot 10^{-6}$ \\
 $\langle J/\psi \rangle / \langle D^0 + {\overline D}^0 \rangle $
&$3.0 \cdot 10^{-4}$ & $4.1 \cdot 10^{-4}$ & $4.6 \cdot 10^{-4}$ \\
 $\Lambda_c$    
&\  2.2   &\  3.7   &\ 6.1    \\
 ${\overline \Lambda}_c$     
&\  0.3   &\  1.7   &\ 4.0    \\
\br
\end{tabular}
\end{indented}
\end{table}

\insertplotg{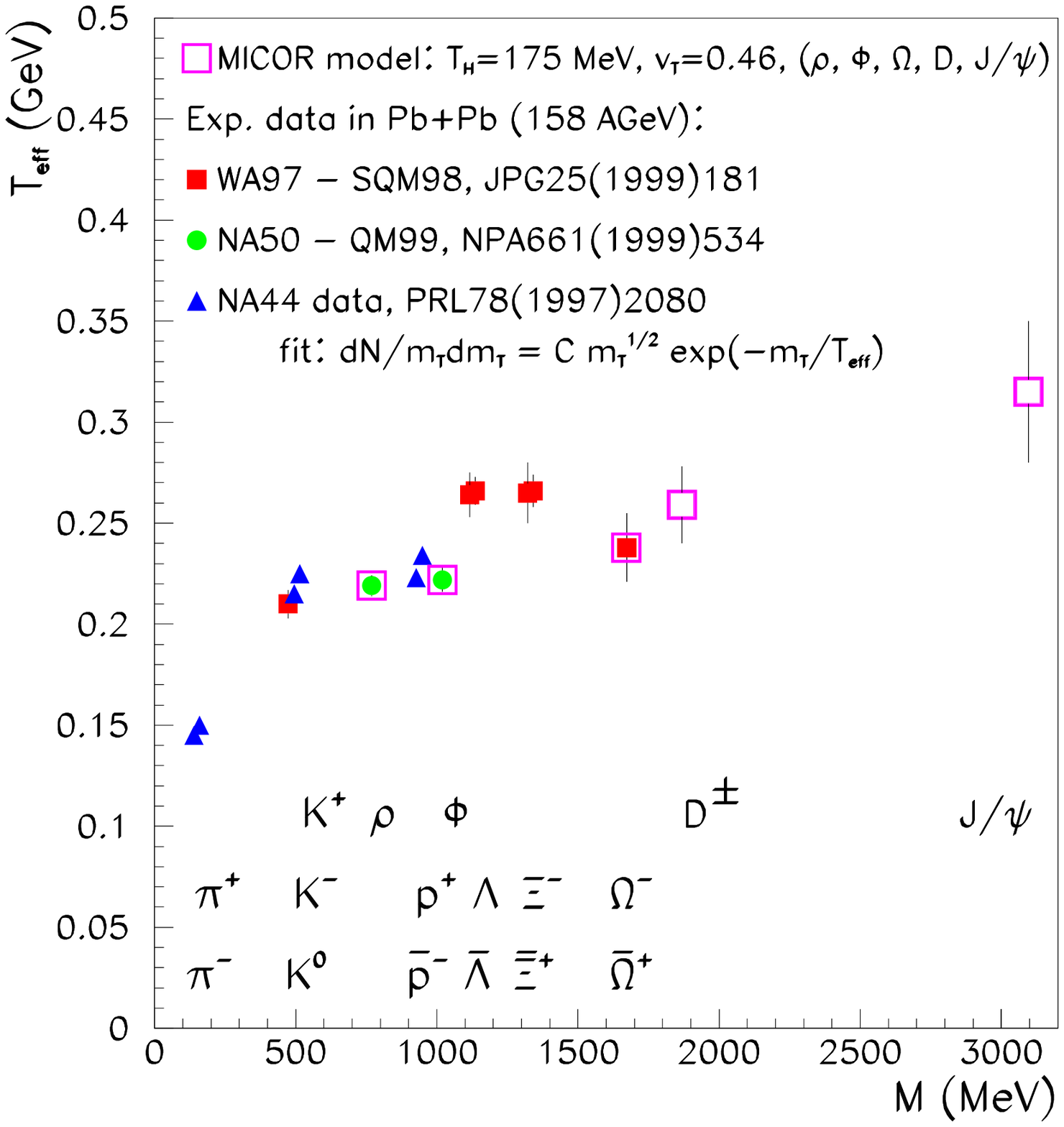}
 \begin{center}
  \begin{minipage}[t]{13.054cm}
  { {\bf Figure 2.}
Experimental hadronic slopes of the transverse momentum spectra
in the PbPb collision at 158 AGeV energy from WA97  (squares),
NA50 (dots) and NA44 Collaborations
(triangulars), see Ref.~\cite{MICORc}.
Open squares with errorbar indicate the MICOR results on 
$D$ and $J/\psi$ mesons. }
  \end{minipage}
 \end{center}

\end{document}